# Noninvasive Blockade of Action Potential by Electromagnetic Induction

Soheil Hashemi [1], Amirhossein Hajiaghajani [2], and Ali Abdolali [1*]

[1] Applied Electromagnetics Laboratory, School of Electrical Engineering, Iran University of Science and Technology, Tehran, 1684613114, Iran

[2] Department of Electrical Engineering and Computer Science, University of California, Irvine, 92697, USA

**Conventional anesthesia methods such as injective anesthetic agents may cause various side effects such as injuries, allergies, and infections. We aim to investigate a noninvasive scheme of an electromagnetic radiator system to block action potential (AP) in neuron fibers. We achieved a high-gradient and unipolar tangential electric field by designing circular geometric coils on an electric rectifier filter layer. An asymmetric sawtooth pulse shape supplied the coils in order to create an effective blockage. The entire setup was placed 5 cm above 50 motor and sensory neurons of the spinal cord. A validated time-domain full-wave analysis code Based on cable model of the neurons and the electric and magnetic potentials is used to simulate and investigate the proposed scheme. We observed action potential blockage on both motor and sensory neurons. In addition, the introduced approach shows promising potential for AP manipulation in the spinal cord.**

## 1. Introduction

Patients who suffer from obesity, diabetes, and potent anticoagulants run the risk of injuring their neurons by traditional methods of general or regional anesthesia [1]. Human sensing receptors communicate with the central neural system by propagating action potentials (AP) [2]. AP blockage of a sensory neuron by static magnetic fields is reported [3]–[5]. However, such a blockade required a long time to emerge and occurred in less than 70% of all AP propagations in the experiments [6].

The external stimulation of neurons by electric (E) fields has been studied in the past decades [7], [8], using the Hodgkin-Huxley cable equations to model neuron stimulation based on the equations derived by the cable theory. In these studies, the E fields that are tangential to the neuron's axis have been considered as the cause of neuron stimulation [9], [10]. Besides, optimization in stimulus pulse shape was conducted, leading researchers into designing an optimum stimulator [11]–[13]. While, we use the theory for investigating the blocking the AP in this paper.

We introduced a new setup that blocks the propagating AP in the neuron fiber and provides anesthesia in the desired parts of the body by inducing a low-frequency tangential E on the spinal cord. This method can instantly affect the nervous system and offers interesting potentials to avoid all the disadvantages of conventional methods including injuries, infections, and other side effects.

## 2. Theory

In order to induce anesthesia in the spinal cord, it is necessary that all neurons block the AP propagation. Incident electromagnetic fields (EM) which block AP along the neuron are likely to reduce the electric potential of the neuron's membrane in one point [8].

To investigate the idea and design the required setup, we have used cable theory. From the modified cable equations mentioned in [8], we obtain

$$\lambda^2 \frac{\partial^2 V_m}{\partial x^2} - \lambda^2 \frac{\partial E_x}{\partial x} - \tau \frac{\partial V_m}{\partial t} - V_m = 0 \quad (1)$$

where $V_m$, $\lambda$ and $\tau$ are respectively the membrane potential, space, and time constants, which are defined in [8]. We appoint the x direction as the neuron axis and refer to the tangential electric fields (E) throughout the present letter by the notation $E_x$. According to (1), $E_x$ must follow $\frac{\partial E_x}{\partial x} < 0$ can reduce the membrane potential. This reduction can cancel propagated AP in the very point if the potential reduction is enough. Creating an EM field with $\frac{\partial E_x}{\partial x} < 0$ in whole space is not possible. Thus, we proposed a spatial $E_x$ formation as shown in Fig. 1a.

---

[*] *Corresponding author: Tel: +98 21 7322 5726, Fax: +98 21 7322 5777; E-mail address: abdolali@iust.ac.ir*

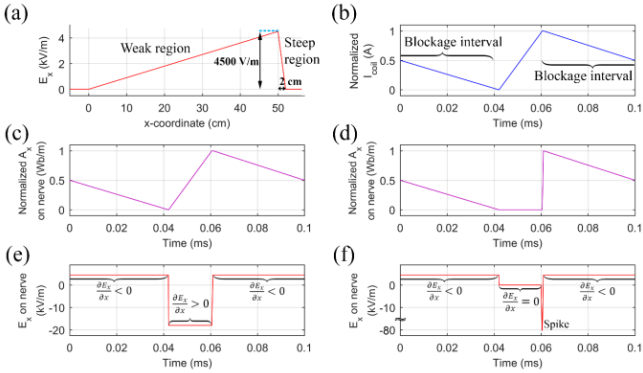

Fig. 1. Theoretical schematic forms of induced tangential E for the AP blockade. (a): Required spatial form according to the cable theory-based equivalent for neuron fiber. (b): Required coil current pulse with shortened degenerative interval and constant slope that gives constant E at its positive cycle. (c,e): Without filtering the degenerative cycle. The steep-decreasing E along the x-axis can block the AP. Nevertheless, the steep-increasing E degenerates the blockade. (d,f): After filtering the degenerative cycle. This is to create a biphasic stimulus pulse with one long weak phase and one brief strong phase. At a moment that diodes turn off, magnetic potential follows the coil currents and rise abruptly. This causes a spike in filtered E. By this filtering procedure, the degenerative interval changes to a spike, whose effect on nerve will be evaluated.

According to (1), the steep descending $E_x$ region (50-52 cm) causes the membrane potential to decrease, such that the AP barrier will be created at the point. On the other hand, an $E_x$ that ascends weakly will not affect the AP significantly. The low-gradient region (0-50 cm) is considered spatially wide in order not to noticeably change the membrane potential. As a result, no AP will be produced or propagated toward the central neural system.

To create this spatial shape of Electric field (E), we used multi-turn coils to radiate an incident EM wave for the production of E along the neuron. Electrical fields is obtained from the Maxwell's equations in general case [14]:

$$\nabla \times \nabla \times \vec{E}(\vec{r}) + \mu\varepsilon(\vec{r}) \frac{\partial^2 \vec{E}(\vec{r})}{\partial t^2} + \mu\sigma(\vec{r}) \frac{\partial \vec{E}(\vec{r})}{\partial t} = -\mu \frac{\partial \vec{J(r')}}{\partial t}$$
(2)

where ε, μ, σ and J respectively represent permittivity, permeability, conductivity and the coil's current density. Also, vectors r and r' denote observation and source coordinates. We have used (2) to design the coils and calculate electric fields.

The coils were placed on the lumbar spine where the skin-muscle tissue is considered as an effective half-space confined by $z < 0$. Frequency should not exceed 10 kHz to avoid thermal side effects in the tissue according to exposure restrictions [15] of E. In 10 kHz, the muscles surrounding a neuron have a relative permittivity, relative permeability, and conductivity of 25909, 1, and 0.34083 S/m, respectively [16], [17].

## 3. Method

We aimed to hyperpolarize the neuron at the blockage point incessantly using an external radiation system. Therefore, the realization of the proposed system revolves around three main phases; first, to appoint $E_x$ as a function of space; second, to filter the degenerative interval in pulse shape; third, to validate the entire setup by the modified full wave analysis achieved from cable model of neurons [18], while exerting the proposed $E_x$.

### 3.1. Developing the spatial form of E

First, the spatial form of $E_x$ must follow the required form for blockade (Fig. 1a). We employed circular coils including 30 turns, placed in the xy plane with a radius of R. A typical coil with the center of $(0,R,z_0)$ exhibits sinusoidal current distribution characteristic in which the x component is very similar to "$\cos\left(\frac{\pi}{2R}x\right)$" for $x \in (-R, +R)$. We observed that, with a very good approximation, the spatial form of $E_x$ (that is itself analogous to the x component of the circular coils' current [8]) and the sine lobes of "$\sin(\frac{\pi}{2R}x)$" are quite similar on the neuron line (y=z+5cm=0). The required $E_x$ which can be denoted by spatial Fourier sine series can be created by a series of circular coils with determined parameters such as radius (relevant to the spatial frequency), current (relevant to the series' coefficient), and coils' center coordinates. Subsequently, a total form of $E_x$ which involves several sine lobes can be created by a setup composed of individual coils that work together (see Fig. 2a). Thus, to form the low-gradient region of the desired $E_x$ (space between 0-50 cm in Fig. 1a), we used three coils (labels 1-3 in Fig. 3).

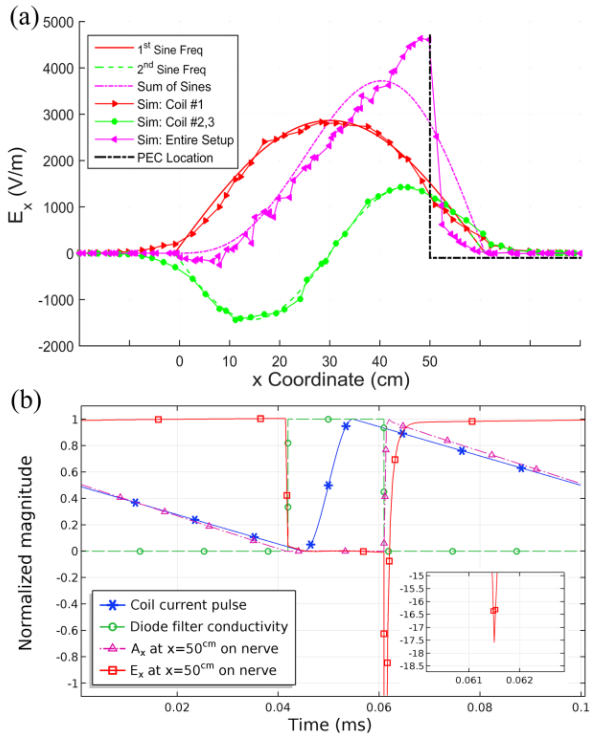

Fig. 2. Results validated by the finite element method. (a) Comparison between analytic sine lobes and simulation of tangential E with PEC plate. (b) Asymmetric sawtooth current pulse is applied to the coils in which the rise time is shortened. The externally controlled diodes filter the E induced only in this interval. Although E remains constant at positive cycles, a spike is engendered by abrupt increase in magnetic potential.

We found a very good agreement between sine functions and simulated $E_x$ depicted in Fig. 2a. Still, to realize the steep region of $E_x$ (space between 50-52 cm in Fig. 1a) without using any additional coil, we employed a thin metallic (perfect electric conductor-PEC like steel) plate under the coils and over the skin. Coils induce E on the plate (label 7 in Fig. 3) instead of neuron, leading to an eddy current distribution.

The PEC dissipates the magnetic field's energy; hence, E drops dramatically where the plate is placed. Also, the high current density on the plate's edge (at x=50 cm) causes $E_x$ to reach its maximum beneath the plate's edge and creates a region with steep $E_x$. Due to IEEE Std C95.6-2002, long term exposure limit of electric fields in very low frequency (VLF) and ultra low frequency (ULF) are 1842 V/m and 10000 V/m consequently though it explicitly mentioned that in a controlled environment it is acceptable to exceed the limit.

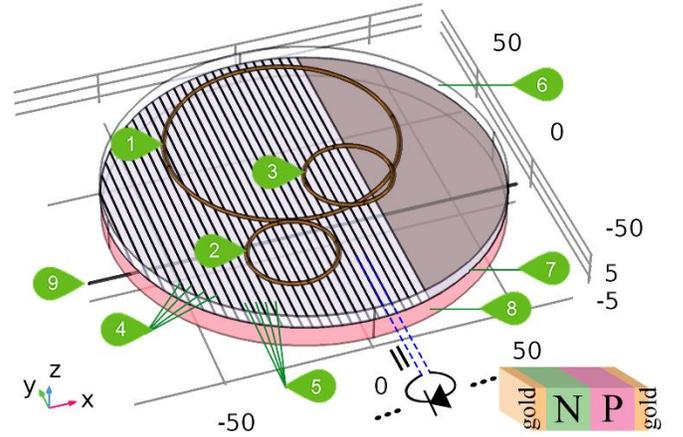

Fig. 3. The entire setup that makes noninvasive anesthesia: 1. Coil No. one; 2. Coil No. two; 3. Coil No. three; 4. Gold rods connected to the external biasing; 5. GaAs layers, equivalent to diodes that work in the x direction; 6. Air; 7. Metallic plate; 8. Effective skin-muscle layer; 9. Neuron axon located 5 cm deep inside the tissue. Coils run currents in clockwise direction.

### 3.2. Filtering degenerative intervals in pulse shape

In addition to (2), from the Biot-Savart law we know:

$$E_x(t) = -\frac{dA_x(J(t))}{dt} \propto -\frac{dI_{coil}(t)}{dt} \quad (3)$$

where A, J and $I_{coil}$ represent magnetic potential, current distribution, and coil ampere-turn, respectively. Inevitably, a periodic pulse for the coils' current includes both rise and fall intervals which, regarding (3), creates both negative and positive $E_x$, respectively. Therefore, it affects the polarity of $\frac{\partial E_x}{\partial x}$. Thus, in order to shorten the degenerative interval, we used a current pulse shape consisting of a long fall time relative to the rise time (see Fig. 1a & b).

It was shown that the ascending interval in the current pulse (Fig. 1b and, consequently, the positive steep $\frac{\partial E_x}{\partial x}$ shown in Fig. 1e) leads to stimulating additional AP since it induces an $E_x$ with negative polarity; hence, E must vanish in this period. We designed a thin filter to be placed beside the metallic plate, involving a repeated sequence of externally biased gold-gallium-arsenide (GaAs) PN junction layers (labels 4 and 5 in Fig. 3). Physically, this means placing many thin diodes whose on/off state is controlled by external biasing. The external biasing of gold layers enables our filter to turn the diodes off in the current's fall time interval (which was creating positive $E_x$ polarity). To specify the method of controlling the induced $E_x$'s polarity, it should be noted that all coils were supplied by synced in-phase current waveform. When diodes are switched off, they act like a dielectric with zero electric conductivity and an effective relative permittivity of 12.9. Thereby, the filter is transparent to the incident E and is allowed to be induced on the neuron behind the filter (see Fig. 2b). In the current rise time, diodes are switched on and act like a good electric conductor with the conductivity of 1 MS/m. This allows E to be induced on the filter instead of the neurons and produce eddy currents, dissipating the magnetic field's energy. This method allows for filtering $E_x$ as shown in Fig. 1f. It should be noted that the Biot-Savart law mentioned in (3) only can be used in linear media. To obtain E in such a non-linear medium, (2) must be employed.

Although the manufacturing considerations of this diode filter have to be well considered and merit additional study, the simulations by COMSOL Multiphysics shows there is an enormous potential for using doped GaAs semiconductors to achieve a conductivity-controllable surface. Other semiconductors with better mobility may be used in the future. It should be mentioned that the thickness of the proposed gold-PN junction is determined such that the spatial form of $E_x$ does not change significantly and the incident wave would not bias the diodes. Table 1 presents the setup parameters.

Table 1. Specification of the final setup. All coils consist of 30 turns.

| Coil No. | Radius [cm] | Current (CW) [kA] | x [cm] | y [cm] | z [cm] |
|---|---|---|---|---|---|
| 1 | 40 | 80.8 | 0 | 30 | 1 |
| 2 | 15 | 69.3 | -15 | -15 | 2.4 |
| 3 | 15 | 69.3 | 15 | 15 | 2.4 |
| Iron dimensions | | | (20,60) | (-60,60) | (0,0.2) |
| Diode layer boundaries | | | (-60,20) | (-60,60) | (0,0.2) |

### 3.3. Validation by the Full Wave Analysis

In the third phase, to investigate our demonstration, we simulated the neuron behavior under the designed setup. As an example of several neurons that were being simulated, a motor neuron was appointed as the neuron fiber inside the spinal cord. We developed the validated full wave analysis including the effect of external stimulus electric fields [8] in Matlab interface using the finite difference method (voltage/time dependencies of ion channels of the membrane are available in [18]). The fiber was exposed to the $E_x$ 0.5 ms after simulation (see Fig. 4). The exposure reduced the membrane's potential at the steep point to create blockade. However, in the weak side, the membrane's potential did not change due to the low gradient of induced $E_x$.

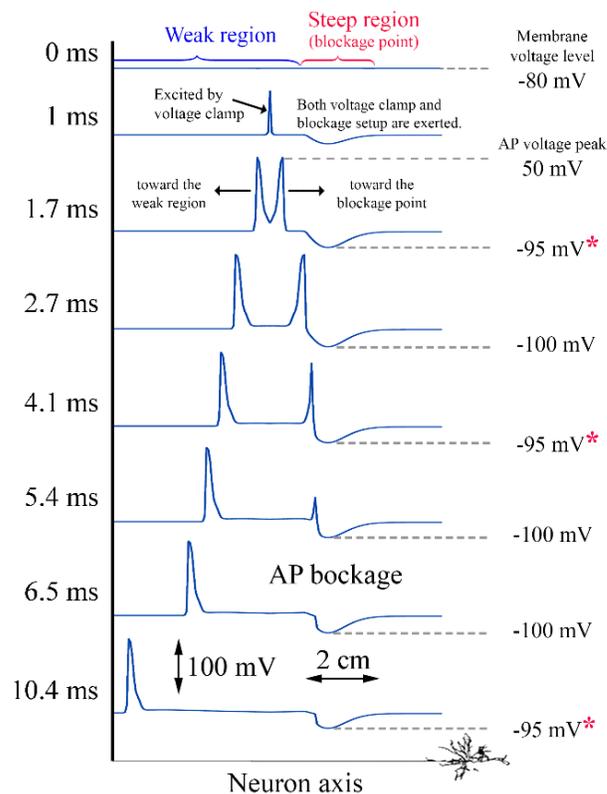

Fig. 4. AP propagation along the motor neuron axis inside the spinal cord. At the beginning, neuron reaches its steady state. Simulation of the blockade setup starts at 0.5 ms. Voltage clamp excites the marked point at 1 ms. Then, the AP propagates toward both ends of the cell, and its blockage can be seen at 6.5 ms when it meets the steep point. The second AP propagates toward the weak region without perceived change. Some of monitored moments in which the electric field's spikes change the membrane potential (at 1.7, 4.1 and 10.4 ms) are starred. It is observed that the negative spikes in Ex result in membrane potential to increase by 5 mV. This voltage fluctuation does not affect the blockage.

The fiber under exposure was excited by a voltage clamp [19] 1 ms after starting the simulation. Afterwards, the AP propagated toward the blockage point (steep region of $E_x$) and the other side of the neuron (weak region). No interference was observed in the membrane voltage while the AP propagated toward the weak region of the induced field. In this case, the neuron was hyperpolarized and went back to rest immediately. However, we directly enforced an artificial incessant hyperpolarization [20] by radiating external E fields such that the AP that was reaching the blockage point could not stimulate the blockage area and the propagation stopped in this region. Because at the point, hyperpolarization avoided neuron stimulation, no AP could continue propagating and consequently, anesthesia was induced. It was observed that neuron remained hyperpolarized despite the short intervals that the diode filter eliminated E (Fig. 1f). As we validated the AP blockage on motor and sensory neurons, we claim that the blockade region can be developed by the proposed setup on every fiber in the spinal cord, causing the AP to stop moving from signal receptors toward the central neural system or vice versa.

### 4. Conclusion

We reported a new form of electric field as a function of space and time, to cancel the AP that was propagating toward the central neural system and induce anesthesia with the use of full wave analysis of neurons. Afterwards, we described a new radiative setup in order to realize the proposed $E_x$ on the neuron fibers located 5 cm deep inside the tissue close by the spinal cord. The setup consisted of three multiturn coils which were controlled by the signal generator, a metallic plate that created the steep $E_x$, and a conductivity-controlled surface to convert the bipolar E to unipolar on neurons. Finally, the validated full wave analysis based on the cable model of neurons was used to successfully investigate and validate the operation of the entire system. Unlike conventional anesthesia methods, this technique eliminates all injuries, contagion, and side effects; it also takes effect immediately after exposure. Therefore, we hope the future researches would illustrate remarkable potentialities of this non-invasive technique by experimenting the proposed setup.